# Celebrating the birth of De Donder's chemical affinity (1922-2022): from the uncompensated heat to his *Ave Maria*

Alessio Rocci[1,2]

[1]Theoretische Natuurkunde VUB and the International Solvay Institutes;

Applied Physics research group (APHY) VUB;

Pleinlaan 2, B-1050 Brussels, Belgium; alessio.rocci@vub.be

[2]University of Padova, Padova, Italy; a_rocci@hotmail.com

https://orcid.org/0000-0002-8943-3544

> Un chimiste qui ne s'intéresserait pas à l'affinité
> Serait comme un poète qui ne s'interésserait pas… à l'affinité!
> (*A chemist who is not interested in affinity is like a poet who is not interested… in affinity!*)
> Aphorisms of Théophile De Donder (recollected around 1927–1930 by Jean Bosquet)

**Abstract**

Théophile De Donder, a Belgian mathematician born in Brussels, elaborated two important ideas that created a bridge between thermodynamics and chemical kinetics. He invented the concept of the degree of advancement of a reaction, and, in 1922, he provided a precise mathematical form to the already known chemical affinity by translating Clausius's uncompensated heat into formal language. These concepts merge in an important inequality that was the starting point for the formalization of the out-of-equilibrium thermodynamics. The present article aims to reconstruct how De Donder elaborated his ideas and how he developed them by exploring his teaching activity and its connection with his scientific production. Furthermore, it emphasizes the role played by the discussions with his disciples who became his collaborators. The paper analyzes De Donder's efforts in participating in the second Solvay Chemistry Council in 1925 to call the attention of the international community of chemists. Even if his mathematical approach did not receive much attention at the time, his work on chemical affinity was the basis for the birth of the so-called *Brussels school of thermodynamics*.

**1 Introduction**

Théophile De Donder (Fig. 1), a Belgian mathematician born in Schaerbeek (Brussels) on the 19th of August 1872, formalized two important ideas in the context of thermodynamics and chemical kinetics. He invented the concept of the degree of advancement of a reaction, also known as extent of reaction, which is a mathematical tool for describing the velocity of a chemical reaction, and he established a precise mathematical form for the already known chemical affinity.

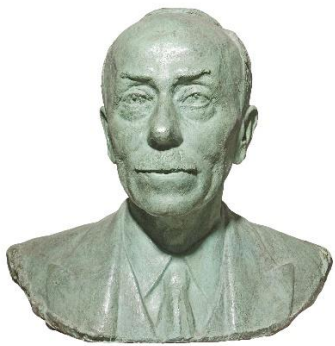

**Fig. 1** De Donder's bronze bust realized by Arthur Dupon in 1957 is on display at the Palais des Académies (Brussels). © Académie royale de Belgique – Luc Schrobiltgen

Michelle Sadoun-Goupil emphasized that "De Donder, with the help of P. Van Rysselberghe, realized the dream of chemical physicists of the end of 17th century"[1] (Sadoun-Goupil 1991, p. 304).

The first aim of this paper is to reconstruct the path that guided De Donder to a mathematical formulation of the concept of chemical affinity, giving it the form accepted today, on the occasion of the centennial of his first paper on this topic (De Donder 1922). The degree of advancement of a chemical reaction with chemical affinity participates in an inequality which, as discovered by De Donder, is a direct consequence of the second law of thermodynamics. This inequality became a pillar on which De Donder's disciples constructed the theory of non-equilibrium thermodynamics. From De Donder's point of view, this inequality was one of the most important outcomes of his approach. The paper's second aim is thus to show how this inequality emerged and to explain the reasons why De Donder attached such importance to this result. By investigating these two aspects, in the present paper we address the following questions: how was De Donder's work received by the community of chemists? How did De Donder try to call attention to his approach? To answer these questions we describe De Donder's attempts to interact with the international community of chemists at the second Solvay Chemistry Council (1925). Independently of these efforts, De Donder's work on chemical affinity proved important with regard to the University of Brussels insofar as it served as a basis for the creation of a Belgian research community, the so-called *Brussels school of thermodynamics* (Kondepudi and Prigogine 2015, p. xxi). The history of this school has not been already analyzed in the literature and it falls beyond the scope of this paper. Despite this, we shall see that De Donder benefited the discussions with his disciples and that they would become his close collaborators. Hence, this paper aims to set the basis for future investigations in this direction.

---

[1] Most of the existing literature on the history of De Donder's chemical affinity is in French. All of the quotations have been translated by the author.

How did De Donder conceive his formalization of chemical affinity? Before De Donder, the most influential position was that of Marcelin Berthelot, who believed that the heat exchanged in a chemical reaction should give a measure of the affinity of the reactants (Sadoun-Goupil 1991, p. 277; Cattani 2010, p. 138). De Donder changed this perspective: as is well known, his starting point was Rudolf Clausius's concept of uncompensated heat.? In this paper, we address the question of why De Donder focused on this quantity by investigating both published and unpublished sources.

What is less well known is that De Donder, trained as a mathematician, delved into thermodynamics because of his first teaching commitment at the University of Brussels (Université Libre de Bruxelles, ULB)[2]. De Donder trained his collaborators, who were students at the ULB. They became professors both at the ULB and all around the Belgium and continued to collaborate with him.

In section 2, we analyse De Donder's scientific production as well as his lectures on thermodynamics until 1920. This enables us to understand how De Donder traced the path that led to the formalization of chemical affinity. De Donder was not the first to apply methods of mathematical physics to thermodynamics and chemistry. He followed the work of Josiah Willard Gibbs and Pierre Duhem, trying to improve them. According to his disciples, in particular Jules Géhéniau, De Donder took a decisive step toward the mathematical formalization of physical chemistry. Quoting Géhéniau: "It may be said that […] De Donder created Mathematical Chemistry […] His attitude towards chemistry was to seek to mathematize it." (Géhéniau 1968, p. 180). What was De Donder's starting point? As we shall see, his impetus came from a lack of theoretical descriptions of irreversible processes, and from his inclination, as a mathematician, to fill this gap. How did his ideas evolve? The present paper shows that De Donder's definition of affinity resulted from an attempt to understand the behaviour of a system as it undergoes an out-of-equilibrium process. Section 3 aims to understand how De Donder identified affinity with the concept of uncompensated heat. In section 4, we analyse the evolution of these mathematical concepts in conjunction with his efforts in calling the attention of the international chemical community to the importance of affinity on the occasion of the second Solvay Chemistry Council, held in Brussels in 1925. In section 5, we briefly analyze when and how De Donder obtained the already mentioned inequality, which paved the way for the modern formulation of the thermodynamics of dissipative open systems (Langyels 1989), and we explain why it was so important to him. Finally, in section 6, we summarize the answers to some questions formulated on the occasion of the centennial of modern chemical affinity.

## 2 The uncompensated heat and the second law of thermodynamics

Robert Raffa and Ronald Tallarida wrote "*Affinity* is familiar to chemists and pharmacologists. It is used to indicate the qualitative concept of *attraction*" [emphasis added] (Raffa and Tallarida 2010, p. 7). By reviewing the history of the use of the word "affinity", the authors underlined how this vague concept became more precise after the introduction in 1875 by Josiah W. Gibbs of a thermodynamical potential, known as Gibbs free energy[3] (Jensen 2015, p. 48). However, Raffa and Tallarida emphasized that it would have been preferable "to have a way of indicating the change in free energy as a function of the extent of reaction, or in other words, the equivalent of the long-sought *driving force*." (Raffa and Tallarida 2010, p. 12). De Donder was attached to this romantic idea of attraction, proposed by Tolman Bergman and popularized by Johann Wolfgang von Goethe in the 19th century. In Goethe's novel *Elective Affinities*, (in German: *Die Wahlverwandtschaften*), the characters discusses Bergman's idea that each substance has its own particular affinity, a sort of driving force acting between chemical elements that can break existing bonds to induce the formation of new compounds. Goethe romantisized this idea suggesting that it can also be used to to describe the shifting relationships of the protagonists of his novel. De Donder, with his work in mathematical physics, found the formal connection between the concepts of force and chemical affinity. The link between De Donder and Goethe is clarified by De Donder's disciple Ilya Prigogine: "Chemistry, portrayed in Goethe's «Elective Affinities» (*of which De Donder was a great reader*), chemistry to which physicists had never really been able to really answer, and the modern enigma of irreversibility thus joined in a challenge that was henceforth unavoidable."[4] (Prigogine and Stengers 1986, p. 207).

---

[2] Until the end of 1960s, there was only one University in Brussels, which then split into two different Universities, i.e. the Université Libre de Bruxelles (ULB) and the Vrije Universiteit Brussel (VUB).

Graduating at the age of nineteen[5], De Donder started his career as a high school teacher and only subsequently pursued higher education in physics and mathematics. He earned his PhD in mathematics from the ULB on November 6, 1899, with a thesis on integral invariants. But why did this mathematician choose to investigate chemical thermodynamics? De Donder started his university career in 1911. His appointment at the ULB (ULB 1911) resulted from a recommendation by Henri Poincaré (Glansdorff et al. 1987, p. 13; Poincaré undated), see Fig. 2. De Donder has attended Poincaré's lectures in Paris, where De Donder has spent one year with the help of a fellowship offered by the Belgian government. Poincaré's lectures has explicitly inspired De Donder's first work (De Donder 1901). As emphasized by Franklin Lambert, retired professor of the University of Brussels, "no recommendation at the ULB in 1911 could be stronger than one from Poincaré, especially for a candidate professor in mathematical physics. Poincaré's figure had an exceptional influence in Brussels: an excerpt of his speech pronounced in 1909 at the celebration of the University's 75th anniversary became ULB's motto. Since then, all professors at the ULB were supposed to subscribe to Poincaré's declaration"[6].

At the time, the physics course was divided into two parts. The first one, taught by Emile Henriot, was a course on general physics. De Donder would teach the second part: a course on mathematical physics. The two most important topics in (classical) mathematical physics at the time were thermodynamics and the theory of electricity and magnetism (Van den Dungen 1958). De Donder's lessons covered both arguments: his lectures on thermodynamics were based on the work of Clausius, Duhem and Poincaré.

---

[3] The concept of affinity appeared explicitly in chemistry at the beginning of the 17th century. The history of this concept is beyond the scope of our paper. For some landmarks see (Jensen 2015, p. 105), while for a more extended treatment see (Sadoun-Goupil 1991).

[4] This anecdote on De Donder is contained in a footnote which is present only in the original French version of the book.

[5] All of the biographical notes are provided by (Van den Dungen 1958) and (Glansdorff et al. 1987).

[6] Private discussion with prof. Lambert. Poincaré's motto is "Thought must never submit to dogma, to a party, to a passion, to an interest, to a preconception, or to anything other than facts themselves; for if thought were to submit, it would cease to be" (Poincaré 1909).

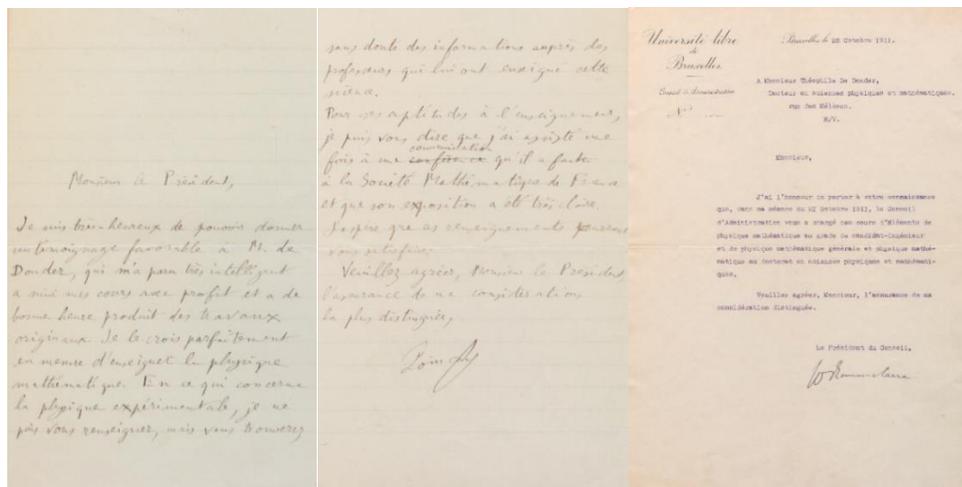

**Fig. 2** On the left and in the center is the (undated) Poincaré's letter; on the right is the letter of the University of Brussels (ULB), specifying De Donder's teaching duties.

De Donder's course reflected his passion for the methods of mathematical physics. These were his words as he ended his opening lecture on November 13, 1911: "Mathematical physics, which is based on pure mathematics, provides us with the purest and the most vivid picture of Nature […] mathematical physics is an inexhaustible source of joy, which makes us love Nature. Its charm nourishes and elevates our souls." (De Donder 1913, p. 11). To explain the methods of mathematical physics, he used the following analogy with chemical reactions: "Mathematical physicists act like chemists who want to reconstruct a material body with the help of the minimum amount of *simplest* elements; in this analogy, the simplest elements would correspond to axioms, the chemical synthesis to the logical arguments; we would have new elements on the one hand, and new phenomena on the other hand, the latter being governed by the laws of mathematical physics" (De Donder 1913, p. 4). In reviewing the various areas of physics, De Donder focused on thermodynamics. On one side, he underlined that the concept of entropy was introduced in the study of *reversible* processes and noticed that the research in this area "made it possible to considerably extend its [i.e. Thermodynamics's] field of application (battery theory, physical chemistry, general chemistry)." (De Donder 1913, p. 8). On the other side, he emphasized how complicated it could be to describe *irreversible* processes from a mechanical point of view. Hence, from the beginning of his career, it was important to De Donder to emphasize the contrast between these two physical processes. In the final part of his opening lecture, after having briefly summarized the last frontiers of physics at the time, e.g. X-rays and Max Planck's quanta of energy, he made the following statement: "All efforts tend to achieve unity: one day, it will perhaps be possible to integrate the Newtonian law of gravitation, the molecular attraction and chemical affinities into the theory of electrons" (De Donder 1913, p. 10). De Donder's speech captured the whole of his research interests. With hindsight, one could interpret his last statement as a sort of research program. Indeed, De Donder spent his entire life investigating Einstein's theory of gravity, chemical affinity, and wave mechanics, trying to apply general relativity in the context of microscopic phenomena.

The key concepts that led De Donder to his mathematical formulation of affinity $\mathcal{A}$ are the uncompensated heat[7] $Q^n$ and the extent of reaction ξ. These three quantities are related by the formula $\mathcal{A} = \frac{dQ^n}{d\xi}$, according to which the affinity is the derivative of the uncompensated heat with respect to the single variable ξ. Our purpose in this section to show how the first of these two concepts entered De Donder's vocabulary of thermodynamics. In his course on mathematical physics, the Belgian mathematician introduced Rudolf Clausius's work and discussed Clausius's idea of uncompensated heat in connection with irreversible processes. Let us consider Clausius's point of view. In his discussion of the form of the second law of thermodynamics, Clausius asserted that the principle may be also expressed as follows: "*an uncompensated transmission of heat from a colder to a warmer body can never occur.*" (Clausius 1867, p. 118; in the footnote). Two important facts must be pointed out. First, Clausius developed the idea of uncompensated transformations, i.e. transformations which involve some real and therefore irreversible processes, but he never introduced an explicit expression for $Q^n$. Clausius introduced a quantity which, like $Q^n$, is a never decreasing quantity but that would represent the entropy produced by irreversible processes (Kondepudi and Prigogine 2015, p. 106). Prigogine and Paul Glansdorff, disciples of De Donder, had noted in an essay dedicated to their master's memory that the uncompensated heat had been briefly studied previously by Lord Rayleigh and Duhem (Prigogine and Glansdorff 1973, p. 680). In this section, we argue that the novelty of De Donder's treatment was its focus on the relationship between $Q^n$ and the extent of reaction[8]. The second point to be underlined is the fact that Clausius's statement of the second law of thermodynamics gave rise to criticisms. Having started his formulation by stating that "heat cannot, of itself, pass from a colder to a hotter body", Clausius realized that the meaning of the expression *of itself* had to be specified: "The words *of itself*, here used for the sake of brevity, require, to be completely understood, a further explanation, *as given in various parts of the author's papers*" [emphasis added] (Clausius 1879, p. 78). Clausius clarified: "heat may be carried over from a colder into a hotter body […] there must either take place an opposite passage of heat from a hotter to a colder body […] This simultaneous passage of heat in the opposite direction, […] is then to be treated as a *compensation* […] we may replace the words *of itself* by *without compensation*" (Clausius 1879, p. 78). In this context, Clausius did not give a mathematical form to his compensation mechanism. The idea therefore remained vague until De Donder's publication of his *Leçons de Termodynamique et de Chimie Physique (première part)* (De Donder 1920c), hereafter called *Lectures I*, which were based on the French translation of Clausius's work, by Françoise Folie, a Belgian astronomer, and Émile Ronkar, a Belgian engineer. In *Lectures I*, De Donder introduced a symbol for the concept of uncompensated heat.

De Donder's *Lectures I* is the published text of two draft versions (Fig. 3), that are archived at the Université Libre de Bruxelles (ULB). The first draft (De Donder 1911) was the text that used by De Donder for the academic year 1911-1912. The second draft (De Donder 1920a) is an improvement of the first draft and it is identical to *Lectures I* in some of its parts. We don't know if it replaced the first draft in classes or if it was a personal copy used by De Donder to prepare *Lectures I*. In the first draft, De Donder expressed his interest in irreversible processes. He noticed the asymmetry between the description of reversible and irreversible transformations, which resulted in the need to enunciate two different versions of the same theorem, one for the reversible and one for the irreversible processes.

---

[7] The superscript $n$ in $Q^n$ means "non-compensée". We preferred the modern term *uncompensated* Instead of the older term *non-compensated*, used e.g. in (Lengyels 1989, p. 445).

[8] The history of the development of the extent of reaction has not been reconstructed to date. This analysis is beyond the scope of the present paper, however.

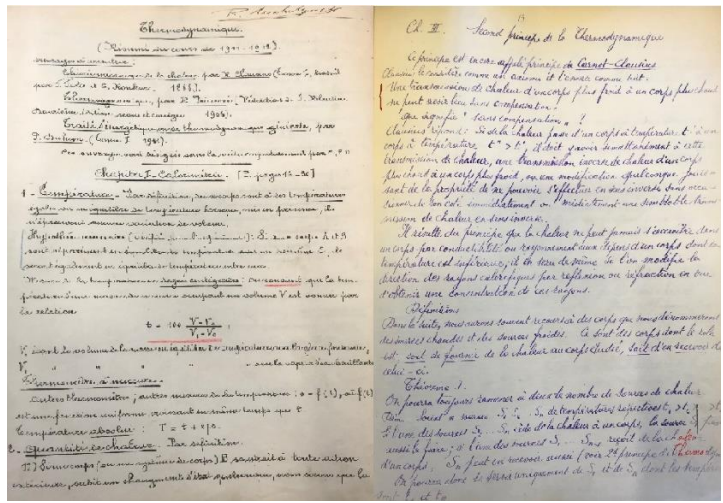

**Fig. 3** The first pages of the two draft versions of *Lectures I* kept in custody by the ULB Archives. On the left the oldest, dated 1911, and on the right, the newest, dated 1920

This fact forced De Donder to extend the theoretical description to incorporate the two processes into a unifying framework. In his draft versions of *Lectures I*, De Donder did not introduce the idea of uncompensated heat, presumably because of the criticisms it had raised. Indeed, reflecting on Clausius's formulation of the second law, he pointed out: "What does [Clausius] mean [with the words] it cannot happen *without compensation*?" (De Donder 1911, p. 8). Thus, having repeated Clausius's statement, De Donder declared: "Clausius's statement is not sufficiently clear; it has been severely criticized." (De Donder 1911, p. 8). He then presented Poincaré's version of the principle: given a system composed of $n$ subsystems, namely $A_1, ..., A_n$, with $T_1 > T_2$, suppose that $A_1$ and $A_2$ undergoes the same transformations, while for the other $n - 2$ subsystems the conditions remain unchanged. If the volumes of $A_1$ and $A_2$ remain unchanged, it is impossible for $A_1$ to become hotter and for $A_2$ to become colder. De Donder expressed his satisfaction by saying: "such must be the formulation of Clausius's principle if one wishes to preserve it from any objection" (De Donder 1911, p. 8). In his later draft, dated 1920 (De Donder 1920a), De Donder repeated Clausius's statement, which specified the existence of a heat exchange that would compensate for the heat transferred from a colder to a hotter point of a system, but did not mention his criticism. Without quoting Poincaré, he presented his argument with the following remark: "It follows from this [Clausius's] principle that the heat in a body cannot be increased, by conduction or radiation, with the heat coming from another body at lower temperature" (De Donder 1920a)[9]. In this later draft, dated to the same year as his *Lectures I*, the second law of thermodynamics is formulated without the introduction of the concept of uncompensated heat. However, the exact quotation is present in De Donder's introduction of the first law in *Lectures I* (De Donder 1920c, p. 36). It can be therefore assumed that De Donder introduced the latter concept during a revision of his lectures.

---

[9] The pages are unnumbered.

When did De Donder decide to revise his draft of *Lectures I* and to publish them? De Donder did not stop teaching his course during the years of the first world war. From 1914 until the armistice of 1918, the University was closed. De Donder gave his lectures at the Athenée Saint Gilles, where he met two students, Frans H. van den Dungen and Georges van Lerberghe, who would help him to revise the draft of his *Lectures I*. We know from the preface that this revised lectures, published in July 1920, resulted from discussions with these students that had started more than one year before. Indeed, during the summer of 1919, in August, van Lerberghe wrote him two letters (Van Lerberghe 1919), dated August 29 and September 6 (Fig. 4), in which he discussed some details with De Donder, e.g. the Carnot cycle and the concept of entropy for arbitrary (real) systems. From these letters, it emerged that van Lerberghe was preparing a provisional report on the new version of *Lectures I*. Hence, these intense discussions stimulated De Donder to improve the draft versions of *Lectures I*.

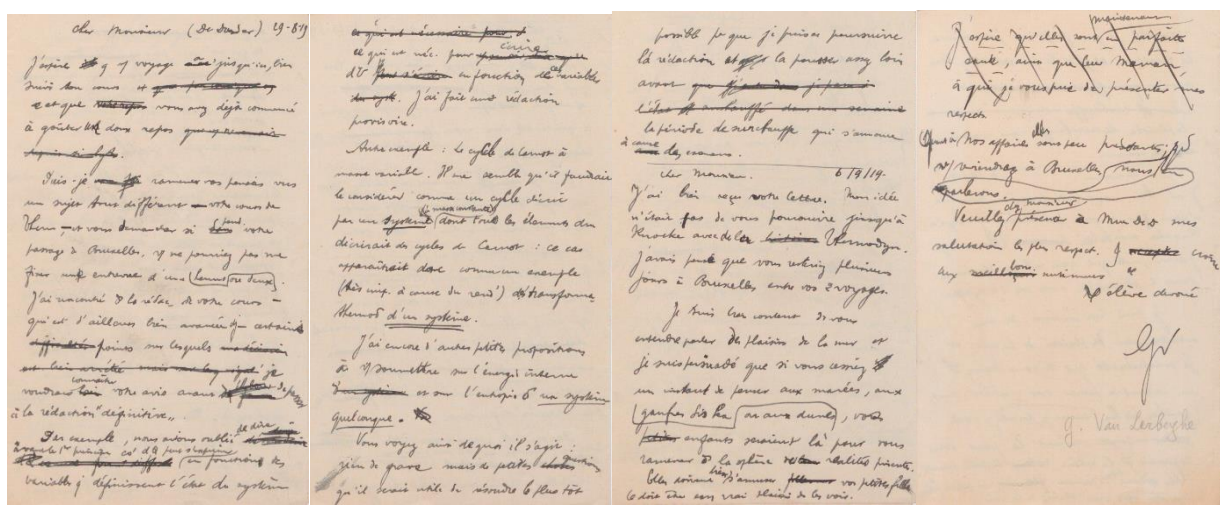

**Fig. 4** Two letters written by van Lerberghe dated 1919 conserved at ULB Archives

The references that appear explicitly in the first draft of *Lectures I* are to the translation of Clausius's *Théorie méchanique de la chaleur* (Clausius 1888), Poincaré's *Thermodinamique* (Poincaré 1908) and Duhem's *Traité d'énergétique* (Duhem 1911). In *Lectures I*, De Donder did not cite Duhem's work, but quoted the recently published book by Max Planck (1913) instead. It is plausible that De Donder had read Planck's work only after the end of the war. Plank emphasized the importance of Clausius's version of the second law: "Only in virtue of this wider meaning of the principle is it possible to draw conclusions about other natural processes." (Planck 1913, p. 85; see the footnote). Hence, Planck pointed out the importance of Clausius's ideas in investigating real phenomena. This was exactly De Donder's focus in his first work on thermodynamics (De Donder 1920b) published after the war but before *Lectures I*. Indeed, in a footnote of the paper, De Donder declared that he had revised his lectures: the paper was published on June 5, while, as already said, the preface of *Lectures I* is dated July 1920. This fact is important because it emphasize the importance played by the revision process in stimulating De Donder's research. In the paper, De Donder underlined that he had started to generalize Willard Gibbs's work to include "real transformations" (De Donder 1920b, p. 316). In the paper's introduction, De Donder emphasized the role of Planck's work: "To our knowledge, there has not yet been a presentation, general, simple and rigorous, of the physical and chemical laws that governs *Gibbs systems*. Among the theorists who obtained some results in this direction we must cite Planck, whose *Thermodynamics* sheds welcome light on large parts of Willard Gibbs's brilliant work" (De Donder 1920b, p. 315). This comment illustrates the impact of Planck's book on De Donder's decision to deepen Clausius's and Gibbs's ideas.

Why did De Donder wait until 1919 to revise his lecture? In De Donder's agenda for the years of war, priority was given to Einstein's theory of gravitation. General relativity was born in November 1915, but De Donder had been aware of Einstein's previous efforts. In his letters to Einstein, he discussed a variational formulation of the theory. De Donder's correspondence with Hendrik Antoon Lorentz, who published his work during the war, confirms the importance of De Donder's role in the development of the variational methods for general relativity. Furthermore, in 1919, De Donder published a series of papers on general relativity in French; a work intended both for communicating his contributions and for attracting attention to the theory in the French-language arena. These preoccupations could explain De Donder's wish to be assisted by van Lerberghe and van den Dungen in his revision of his lectures on thermodynamics. They also allow us to understand why these lectures remained unpublished until 1920.

Why did De Donder decide to introduce a symbol for the uncompensated heat? In the following, we argue that he was motivated by the fact that the new symbol would enable him to clarify the meaning of the term *uncompensated* with the use of precise mathematical language. Comparing the drafts with *Lectures I*, it can be noted that De Donder added a chapter where he analyzed the consequences of both the first and the second law of thermodynamics. In this chapter, he explicitly formalized Clausius's idea: "The inequality [second law] leads us to write $dQ = TdS - dQ^n$, where $dQ^n$ is zero or a positive quantity, called *the uncompensated heat* […] We will justify this definition introduced by Clausius in his investigation of irreversible processes" (De Donder 1920c, p. 80). Hence, De Donder took a step forward when he found a way to translate Clausius's original statement into a mathematical expression. De Donder was guided in his research by the need to formalize real processes, an objective that he would reaffirm in his later works. Thus, he clarified the term *uncompensated* as follows: the quantities $Q_{ACB}$ and $Q_{BCA}$, which represent the heat exchanged during two opposite transformations, cannot have the same absolute value, as is the case with for reversible processes. For a cyclic transformation involving real phenomena, their sum must be a negative quantity. The change in entropy for any process leading to a transformation between an initial state $A$ and a final state $B$ being $S_B - S_A$ must be zero after a cyclic transformation, even if this transformation involves real processes. Hence, it follows from De Donder's formulation of the second principle that $Q_{ACB} + Q_{BCA} = -(Q^n_{ACB} + Q^n_{BCA})$. This quantity is negative because the second law implies both $Q^n_{ACB} > 0$ and $Q^n_{BCA} > 0$ for real processes. This means, in particular, that $Q_{ACB} \neq -Q_{BCA}$. De Donder concluded that "hence, the two heat quantities *do not compensate each other* during the two opposite transformations: this observation explains the meaning of the term *uncompensated heat*." (De Donder 1920, p. 81).

What were the reactions to De Donder's book? Adolphe Buhl, another mathematician, wrote an enthusiastic review. He praised De Donder's formal approach and underlined his "deep and delicate sense of the physical reality" (Buhl, 1921, p. 351). However, it is worth emphasizing that De Donder was more interested in the formal aspects of the theory than in its verifiable consequences. Buhl pointed out that De Donder improved the approach initiated by Gibbs and that De Donder emphasized the importance of the thermodynamic potential like in Duhem's work. Finally, he briefly mentioned the innovative character of De Donder's text, which presented the generalization of Braun's, van 't Hoff's and Le Chatelier's laws in the case of non-constant temperature. As a mathematician, Buhl appreciated the use of calculus and the extensive application of the theory of differential forms with several variables. As we shall see in the following, however, De Donder's approach, which represented a "triumph of the mathematical spirit" applied to chemical processes, was not appreciated at the beginning by chemists for at least three reasons: first, because "many nineteenth-century chemists had minimized the dependence of their field on hypothetical reasoning and stressed its basis in empirical laws" (Nye 1993, p. 110); second, because, as underlined by Bensaude-Vincent and Stengers, "at that time the thermodynamics seemed to be a closed case" (Bensaude-Vincent and Stengers 1993, p. 314); and, finally, because "[the chemists] were not prepared to use the mathematical language introduced by Gibbs" (van Tiggelen 2004, 101) and later developed by Duhem and De Donder.

**3 The birth of modern chemical affinity (1922)**

In 1920, De Donder started to investigate the relationship between thermodynamics and chemistry. As Brigitte van Tiggelen has pointed out, "the contact between chemistry and thermodynamics in the 20th century produced not only a synthesis of the two fields or an extension of the limits of thermodynamics on the domain of chemistry but also a sort of revolution for the thermodynamics itself, which started to address new important characteristics that are fundamental during chemical processes" (van Tiggelen 2004, p. 100). This revolutionary process started with the work of Gibbs and Duhem, who introduced an advanced mathematical language. This was an innovation that chemists of the period were not prepared for, but a challenge that De Donder could face because of his mathematical background. Van Tiggelen noted that "despite the progress in mathematizing the description of chemical processes, it was not possible to deduce the direction of the reactions from the equations" (van Tiggelen 2004, p. 101) and that a key step in this direction was De Donder's introduction of the extent of reaction. According to the chemist Gianni Astarita, this concept was first put forward by De Donder (Astarita 1989, p. 66). De Donder introduced the degree of advancement in 1920 in the first paper where he started to investigate the systems he called *Gibbs systems*, a mathematical abstraction introduced by De Donder to describe all the configurations of a system that undergoes both physical and chemical transformations (De Donder 1920b). Every transformation is called a *phase* of the system and it is characterized by a set of variables. In this paper De Donder was interested in modelling real processes with his Gibbs systems and hence, he specified that the single configuration should not represent only systems at equilibrium. In this context, he needed a parameter, i.e. the extent of reaction, representing the displacement from an initial to a final configuration through the phases of the system.

Both in (De Donder 1920b) and *Lectures I*, this new parameter, defined by the ratio of two quantities, had no name. In (De Donder 1920b), De Donder did not explain the meaning of the ratio, even if it can be inferred by analyzing the sign of the quantities involved. Indeed, he noted that the sign of the ratio is connected to the direction of the chemical reaction modelled by his Gibbs system. In *Lectures I*, where there is a detailed explanation of the arguments presented in (De Donder 1920b), De Donder introduced the symbol $\mu$ for the new parameter and he stated explicitly that its virtual increment $\delta\mu_\rho$, which was defined for every single chemical reaction labelled by $\rho$, would represent the progress of the chemical reaction (De Donder 1920c, p. 118). De Donder used the symbol $\mu$ until 1925, when he changed it to $\xi$. At the end of section 4, it will be clarified why he did not make a mere graphical change.

In the following, we argue that Duhem's book played an important role in shaping De Donder's approach. Even if he did not quote it in *Lectures I*, the definition of the Gibbs systems followed a path traced by Duhem. In the earliest draft version of *Lectures I*, De Donder explicitly followed Duhem's manual, introducing what he called "the generalization of Massieu's first characteristic function" (De Donder 1911, p. 35), i.e. the enthalpy. As René Lefever has stated, "the study of equilibrium stability, [which is] mainly due to Gibbs […], it is based on the thermodynamic potentials method that, as Duhem noted, is only applicable in special equilibrium cases where such potentials exist." (Lefever 2017, p. 2). To get rid of this restriction, Duhem had introduced the generalization of some potentials as follows. Given a thermodynamic potential H(a,b,c) depending on some variables a, b and c, Duhem's generalization is obtained by defining a new potential, now $\mathcal{H}$(a,b,c; A, B, C), which now is a function both of the old variables and some new variables *A, B, C*, and supposing that the old potential can be obtained from the new one by setting to zero the values of the latter. These new auxiliary variables would describe out-of-equilibrium states of the system. The new potential was indicated by Duhem as the true instrument to describe the dynamic of the system (Duhem 1911 p. 253). Duhem worked in analogy with a mechanical system and, in this context, the partial derivative of the new potential with respect to one of these new variables would represent a sort of *internal force*. Neither Duhem nor De Donder used this expression, but they represented the equilibrium condition of the system as a set of equations where every "external force" (De Donder 1920a, p. 35) equals the derivative of the potential with respect to the variable causing the virtual work associated to this force. It is worth noticing that this idea contained *in nuce* the idea of affinity, which indeed is the derivative of Gibbs's free energy $G$ with respect to the extent of reaction, which in turn describes the displacement of the system out of a state of equilibrium. It is through its characterization in terms of $G$ that the connection between affinity and the notion of attraction (and repulsion), which was due to some vague force in the context of alchemy (Sadoun-Goupil 1991), emerged in De Donder's work. To understand this point, it is important to notice that De Donder used Duhem's generalization process to introduce generalizations of existing concepts. By following Duhem's path in investigating out-of-equilibrium systems, De Donder identified the state of equilibrium with the minimum of the new function he introduced. More precisely, in (De Donder 1920b) and (De Donder 1920c), the thermodynamic potential he considered was Gibbs free energy[10] and the additional variables were the masses $m_\gamma^a$, where $\gamma$ labelled the constituents of the system while the index $a$ indicated the specific phase. To obtain a minimum, the first condition to be fulfilled is the request to have a stationary point for the (modified) Gibbs free energy[11]. Let $\nu_\gamma$ be the stoichiometric coefficients and $M_\gamma$ the molar masses. The condition to have a stationary point reads $\sum_{\gamma,a} \nu_\gamma^a M_\gamma^a \frac{\partial G}{\partial m_\gamma^a} = 0$ (De Donder 1920c, p. 121; De Donder 1920b, p. 318). Two years later, the quantity on the right-hand side would be related to the affinity and the equilibrium would correspond to the case of zero chemical affinity, as De Donder would recognize after having figured out the connection between Gibbs free energy and the uncompensated heat. Before proceeding, it should be noticed that one of the outcomes of the paper dated 1920, also explained in more detail in *Lectures I*, was a formulation of Le Chatelier and van 't Hoff laws involving the degree of advancement of the reaction (De Donder 1920c, p. 122-123; De Donder 1920b, p. 11). Maybe, because of this fact, De Donder sent a copy of *Lectures I* to Henri Le Chatelier (Le Chatelier 1920), see Fig. 5.

---

[10] In (De Donder 1920a), (De Donder 1920b), and (De Donder 1920c), De Donder did not use the letter $G$, which is the modern symbol. In 1920–1922 and in the following years, he used different letters for this thermodynamic potentials. A complete list can be found in (Sadoun-Goupil 1991, p. 305).

[11] De Donder considered also the second condition, which would select the minimum among the stationary points. We shall not delve into this technical details.

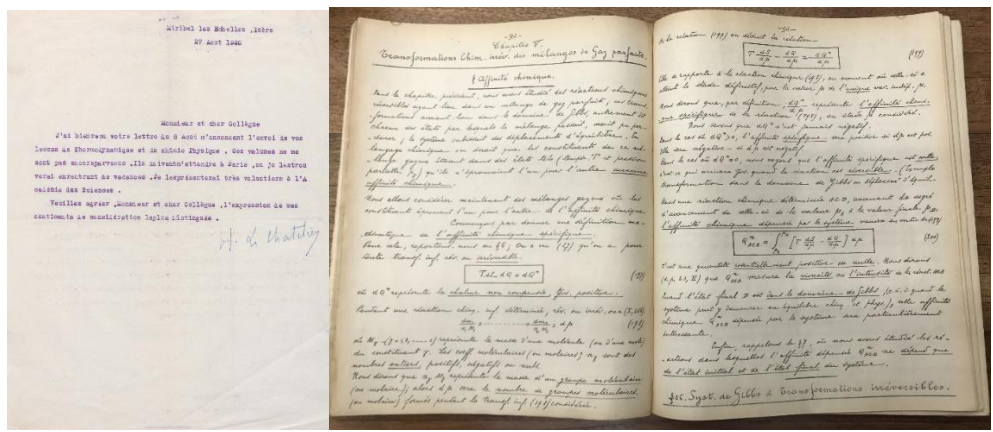

**Fig. 5** On the left, is Le Chatelier's letter. On the right, the fifth chapter of the (unpublished) *Lectures II* on chemical affinity

In 1921–1922, De Donder focused again on general relativity and published both a book, entitled *La Gravifique Einsteinienne*, and supplements of the book. Besides this, he continued to develop his *Lectures I* after their publication. Indeed, as stated in the preface of his *Lectures I*, they were conceived as the first of two parts: the first devoted to the general theory, the second that delved into the applications. This second part, hereafter *Lectures II*, remained unpublished and in an hand-written form. There is no date in the whole document: for this reason, in the references, no date has been indicated and it will be quoted as (De Donder *Lectures II*). The text of *Lectures II* is more than a hundred pages long and it is plausible that the various chapters were developed and written in different years. Even if there are no explicit documents proving this fact, we can present the following arguments. It can be argued that *Lectures II* were developed between 1920 and at least 1922, because the fifth chapter of *Lectures II*, starting on page 92, contains both the definition of chemical affinity and all the topics De Donder briefly investigated in papers published in May and August 1922, but with more detailed explanations and in a more pedagogical form. In the first four chapters, when De Donder explicitly indicated that he would clarify a concept in the subsequent paragraphs, he investigated the topic after one or few pages and he never referred to a following chapter. Moreover, until chapter five he never mentioned the concept of affinity. From our point of view, it is therefore plausible to assume that if he had discovered the connection between uncompensated heat and affinity, he would have published it, or at least mentioned it, before 1922. Indeed, De Donder was used to integrating his latest achievements in his lectures, as he did in *Lectures I*. Furthermore, in the second chapter of *Lectures II*, De Donder presented some partial results without concluding what he would explain using the concept of affinity in the fifth chapter. In chapter two, he showed the connection between Gibbs free energy and the uncompensated heat only for transformations that are both isobaric and isothermal. In the same chapter, by investigating the consequences of this connection for Gibbs systems and discussing irreversible processes, De Donder pointed out that the equality $\sum_{\gamma,a} \nu_\gamma^a M_\gamma^a \frac{\partial G}{\partial m_\gamma^a} = 0$ would become an inequality, namely $\sum_{\gamma,a} \nu_\gamma^a M_\gamma^a \frac{\partial G}{\partial m_\gamma^a} < 0$, but he did not mention how the left-hand side of the inequality is connected to the uncompensated heat as he would point out explicitly both in the first paper on affinity (De Donder 1922a, p. 198) and in the fifth chapter of *Lectures II*. In chapter four, De Donder investigated the relationship between the equilibrium condition $\sum_{\gamma,a} \nu_\gamma^a M_\gamma^a \frac{\partial G}{\partial m_\gamma^a} = 0$ and the Guldberg-Waage law, also known as mass action law, for perfect gases: "an important equation for Theoretical Chemistry and for studying the processes of the modern chemical industry" (De Donder *Lectures II*, p. 71). Indeed, this equation is extensively analyzed in this chapter. De Donder showed explicitly the equivalence between the law and the equilibrium condition: in *Lectures II* his treatment is twenty pages long and he only considered reversible processes. Hence, it is not surprising that, in his first paper on chemical affinity, the first example considered by De Donder for describing irreversible processes would be the generalization of the Guldberg-Waage law (De Donder 1922a, p. 200). In chapter five of *Lectures II*, De Donder derived also van 't Hoff and Le Chatelier laws that he had considered in 1920. Although all the papers written between 1864 and 1879 by Guldberg and Waage contained the concept of affinity, De Donder did not mention this fact in his fourth chapter of *Lectures II*. From our point of view, the missing reference to the concept of affinity is another indication that he did not have realized at the time how the concepts of uncompensated heat and Gibbs free energy can be related to chemical affinity.

After having presented our pieces of evidence that Lectures II covers at least the period between 1920 and 1922, we arrive at the genesis of chemical affinity. Why did De Donder decide to introduce the concept of affinity? As we have already emphasized, De Donder knew the analogy between the concept of affinity and the idea of a driving force. After having explicitly introduced the uncompensated heat and after having understood that, at the equilibrium, no uncompensated heat is exchanged, it is plausible to infer that he supposed that the two quantities should be connected. It is instructive to quote his comments in the fifth chapter of *Lectures II*. It started as follows:

> "In the preceding chapter, we have studied reversible chemical reactions of mixtures of perfect gases; this transformation occurs in Gibbs's domain, i.e. where anyone of the states of the mixture of gases would be a stable one; the transformation proceeds through a sequence of equilibrium states. In chemical language, it can be said that the constituents of the gas mixture are in a state such that (temperature $T$ and partial pressure $p_\gamma$) they cannot feel *any chemical affinity* among each other. Hence, we shall now consider when they feel some chemical affinity." (De Donder *Lectures II*, p. 92)

After this incipit, De Donder recalled how he had introduced the uncompensated heat and the extent of reaction $\mu$, underlining the importance of the former, but without choosing a name for the latter. This variable finally deserved a name in (De Donder 1922a), where De Donder explicitly defined it as "the degree of advancement of the reaction considered" (De Donder 1922a, p. 199). Both in *Lectures II* and (De Donder 1922a), he was finally ready to define what he called the "specific chemical affinity" $\mathcal{A}$ (De Donder *Lectures II*, p. 93) using the two differentials $dQ^n$ and $d\mu$, namely $\mathcal{A} = \frac{dQ^n}{d\mu}$. As already said, he considered Gibbs systems undergoing irreversible processes and he clarified that, out of the equilibrium, for constant values of pressure and temperature, the quantity $\sum_{\gamma,a} \nu_\gamma^a M_\gamma^a \frac{\partial G}{\partial m_\gamma^a}$ is no longer zero but it is negative, since he was considering a minimum. Hence, this quantity can be identified with the opposite of the specific affinity, which is always positive due to the second law. Finally, De Donder obtained the connection between uncompensated heat, affinity and Gibbs free energy, namely $\sum_{\gamma,a} \nu_\gamma^a M_\gamma^a \frac{\partial G}{\partial m_\gamma^a} = -\frac{dQ^n}{d\mu}$. De Donder declared on p. 95 of *Lectures II* that this new equation represents "the generalization of the equation of Guldberg and Waage for irreversible processes". This statement is present also in (De Donder 1922a, p. 4), where, in a footnote, De Donder explicitly promised to clarify it in the forthcoming *Lectures II*. Indeed, the assertion should have sounded obscure at the time because, in *Lectures II*, it required a notable effort to show that when the affinity is zero the equation is equivalent to the mass-action law of Guldberg and Waage.

The Belgian mathematician also published two brief communications (De Donder 1922b), in May and in August of 1922, where he discussed the relationship between Nernst's theorem and the affinity. We shall not enter into details, but it is worth noticing that their content is contained in *Lectures II*. This means that he continued to develop his research and that he regularly systematized it to include all his achievements in his teaching activity[12]. With this practice, he educated the generations of physical chemists that would form the Brussels school of thermodynamics.

**4 The chemical affinity and the second Solvay Chemistry Council (1925)**

Between 1922 and 1925, De Donder did not publish anything on thermodynamics or chemical affinity. He continued to teach his course in mathematical physics, but his efforts were focused again on developing general relativity and disseminating it. De Donder published ten papers on general relativity and he prepared three books[13], namely *La gravifique de Weyl-Eddington-Einstein* published in 1924, his *Théorie Mathématique de l'électricité* in collaboration with van Lerberghe, and his *Introduction a la Gravifique Einsteinienne*, both published at the beginning of 1925. Furthermore, he presented Einstein's theory by lecturing at various Universities. In March 1923, De Donder was invited to the University of Nancy and Strasbourg for a conference on Einstein's theory, while the following year, in May, he was invited for the same reason to the Sorbonne in Paris. His efforts and his pedagogical approach were fully appreciated and in 1924 he was awarded the title *honoris causa doctor* from the University of Nancy. Between the end of 1923 and the end of 1924, De Donder discussed some questions concerning the theory of relativity with Sir Arthur Eddington, as evidenced by the correspondence conserved at ULB. In the same year, he was awarded the *Prix décennal des mathématiques appliquées*, i.e. a decennial prize in applied mathematics for the years 1913-1922, with other three scientists[14]. The review of his work made by the jury[15] for the Ministry of Arts and Sciences will be quoted again in this section. In a letter dated March 23, 1924, Lorentz invited De Donder to attend the fourth Solvay Physics Council, which was held in Brussels between 24-29 April. This was De Donder's official entry into Solvay's meetings. The father of the Councils, Ernest Solvay, was aware of De Donder's work (Fig. 6). Indeed, he received De Donder's work on general relativity in October 1921 (Solvay 1921). But the Belgian mathematician had not yet been actively involved either in the Physics Solvay Councils or in their Chemistry counterparts, which started in 1922, soon before Solvay's death. Also, the year 1925 was a very busy period for De Donder. At the beginning of this year, after the publication of his book on the mathematical theory of electricity in collaboration with van Lerberghe, he sent some copies to his colleagues around the world: Josef Larmor and Ebenezer Cunningham in Cambridge, Sydney Chapman in London, Vito Volterra in Rome, Edward Arthur Milne in Dublin and Lorentz in Netherlands and to libraries, as evidenced by De Donder's correspondence. On the February 17, he sent a copy also to Paul A. Heymans, a Belgian colleague who was Assistant Professor at M.I.T. from 1922 (Fig. 6). In a letter dated March 16 (Heymans 1925), Heymans acknowledged De Donder and asked him for a copy of both a De Donder's paper on a Boltzmann's theorem published in February, and De Donder's paper on affinity. Heymans specified: "I showed your letter to professor Debye, and after discussing the matter together, we are both anxious to receive your views on that subject" (Heymans 1925).

---

[12] The chronology of inclusion in *Lectures II* is unclear because of the lack of explicit dates in *Lectures II*.

[13] For a brief comment of the books see Glansdorff, Bosquet and Géhéniau (1987).

[14] The other scientists were Jules Boulvin, Giuseppe Cesàro and Auguste Collard.

[15] The jury was composed by François Keelhoff (University of Gand); André Duchesne and Henri Buttgenbach (University of Liége); Paul-Henri Stroobant (Brussels University); Gustave Gillon (University of Louvain).

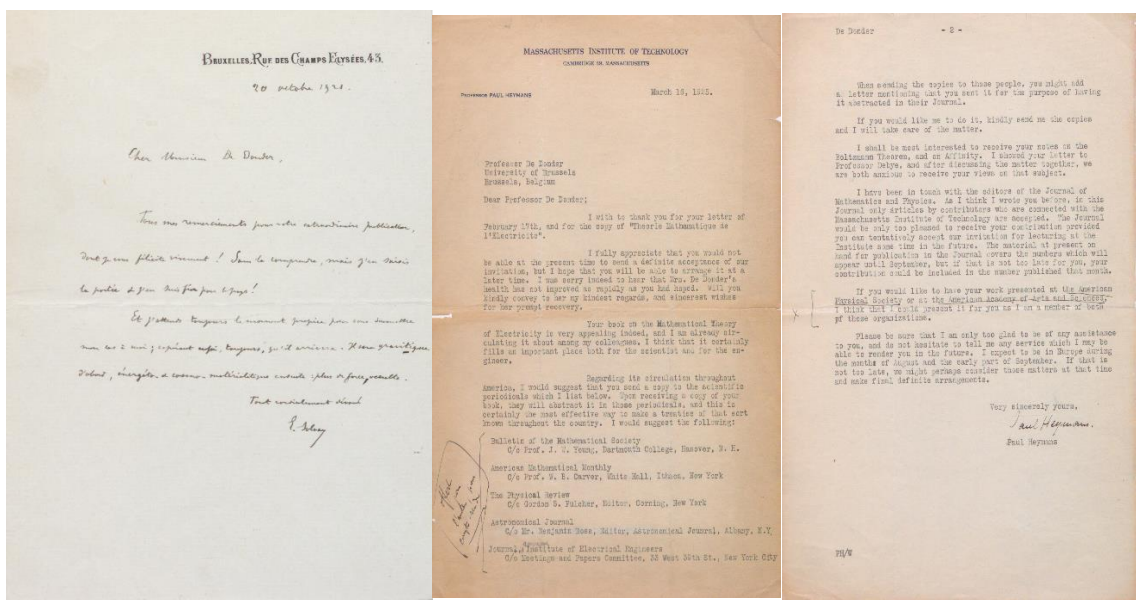

**Fig.6** On the left is Ernest Solvay's letter. The other pictures refer to Heymans' letter

Although chemical affinity was not a priority for De Donder, he continued to discuss it with van Lerberghe. The collaboration between the two authors focused on De Donder's book on the mathematical theory of electricity, but van Lerberghe had started to teach the course of mathematical physics in February 1924 at the University of Mons, where he had also received the funds to create a laboratory for thermodynamics. Van Lerberghe started to investigate the relationship between the affinity and the velocity of direct and inverse reactions, and their correspondence is proof that De Donder was aware of his work. In 1925, De Donder presented again new results on chemical affinity to the Royal Academy, where he regularly participated in meetings (De Donder 1925d; De Donder 1925e; De Donder 1925f). The three papers are dated May, June and August respectively. Before the publication of his papers, he also tried to call the attention of chemists to his work. On March 7, his book on the mathematical theory of electricity appeared in *Hommages d'Ouvrages*, a section of the Bulletin of the Belgian Royal Academy for the Sciences, which usually celebrated newly published works. As already noticed, on March 16, De Donder received Heymans's letter quoted above, see Fig. 6, who recalled his attention to chemical affinity. Soon after, on April 4, at the monthly meeting of the Royal Academy, *Hommages d'Ouvrages* celebrated the new volume of the proceedings of the first Chemistry Solvay Conference. At this monthly meeting, both De Donder and Octave Dony-Hénault, the secretary of the scientific committee for the Chemistry Councils, were present. Hence, it seems plausible that De Donder and Dony-Hénault talked about the forthcoming second Chemistry Council during this meeting. Indeed, as it emerges from a letter dated April 9 written by Dony-Hénault to Charles Lefebure, the secretary of the administrative commission, De Donder asked him to refer his request to the scientific committee to read a communication about "his studies in thermodynamics on the problem of affinity" (Dony-Hénault 1925) at the Council (Fig. 7).

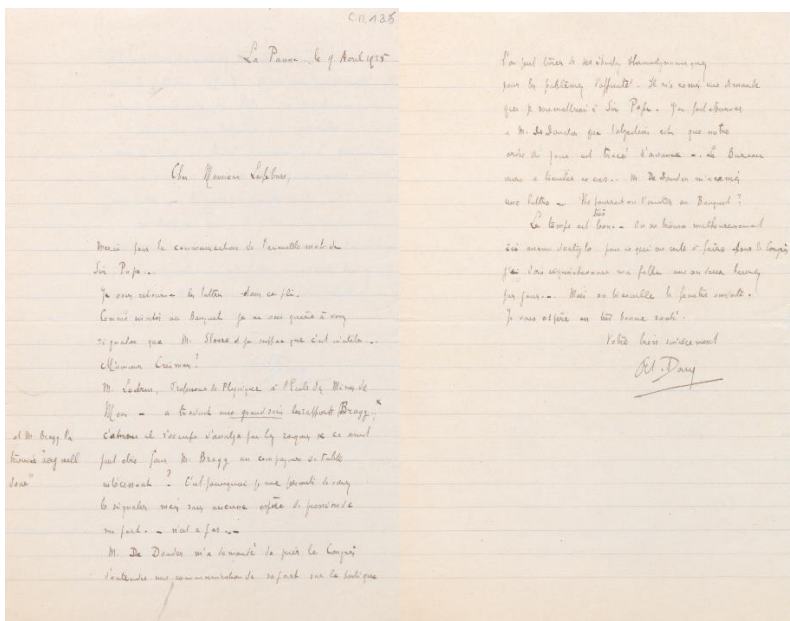

**Fig. 7** Letter of Dony-Hénault to Charles Lefebure (Solvay Archives)

In the same letter, Dony-Hénault said that he would refer the proposal to William Pope, the president of the scientific committee, although he warned De Donder that the program had been scheduled in advance. As a solution, in the same letter, Dony-Hénault proposed to Lefebure to invite De Donder to the official banquet offered by Armand Solvay[16] on April 22. In all the preceding Solvay Councils, the Physics and the first Chemistry conferences, the banquets were not only a social event but also an occasion where topics connected with the conferences were debated. Hence, this was the reason for the official invitation. In Fig. 8, two documents are shown where, on the front page, there are two handwriting annotations. By using the Solvay Archive of ULB, it emerged that De Donder gave these two documents to Lefebure. Indeed, as shown in Fig. 8, on the left, Lefebure annotated in pencil De Donder's name. Lefebure's calligraphy has been checked with the various letters contained in the Archive. In Fig. 8, on the right, there is De Donder's signature, which can be confronted with his signature in (Glansdorff et al. 1987, see p. 4)., see Fig. 8.

---

[16] Armand was the son of Ernest Solvay, who was died in 1922.

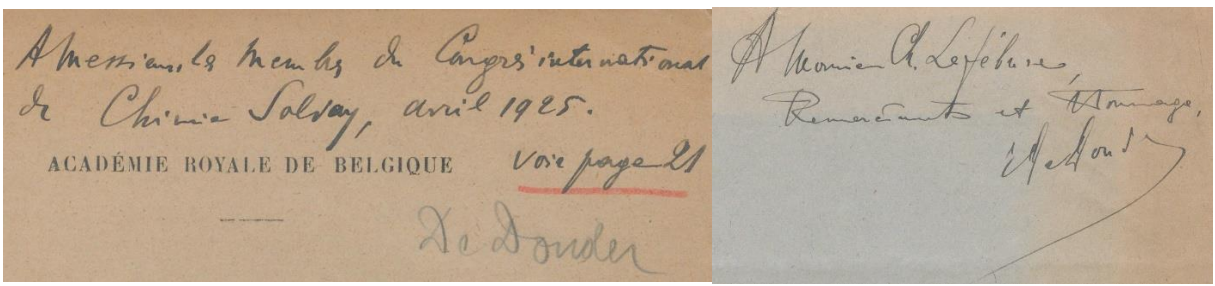

**Fig.8** The handwritten annotations on the first page of the two documents donated by De Donder to Lefebure. On the left is the extract of the Bullettins. The name *De Donder* in pencil is Lefebure calligraphy. On the right is the copy of (De Donder 1922) with De Donder's signature

One document was the extract of the *Bullettins de la Classe des Sceances* (De Donder 1925a) evaluating De Donder's scientific activity and announcing that he received the decennial prize in 1924, maybe meant to emphasize the importance of his achievements. On the front page De Donder wrote "À Messieurs les membres du congrès international de Chimie Solvay, avril 1925. voir page 21"[17]. Inside the document, the parts of the text where the report described De Donder's activity on thermodynamics and chemical affinity are underlined with a red colour like on the front page, shown in Fig. 8. The second document is a copy of his first work on chemical affinity (De Donder 1922a). The dedication in the first two lines of the front page reads "À Monsieur Charles Lefébure. Remerciements et hommages"[18]. Both handwriting and the signature can also be compared with the calligraphy of De Donder's letter, which we quote in the following (Fig. 11). We recognized again De Donder's signature and, as it emerges clearly by comparing the various documents, the handwritings belong to the same author. Finally, De Donder prepared many copies of an unpublished communication[19] (De Donder 1925b), entitled *Affinité chimique et Vitesse de Réaction*, concerning chemical affinity and the velocity of reactions, hereafter the *Solvay communication*.

The *Solvay communication* was conceived both to present De Donder's ideas to the international community of chemists and to discuss the connection between his work and the problems that emerged in 1922 during the first Chemistry Councils. In the following, we shall analyze these two aspects. At the beginning of the communication, De Donder presented his focus and his goals: "In a note published in the Bulletins de l'Académie Royale de Belgique (Classe de Sceance) on 2 May 1922 [namely (De Donder 1922a)], we have identified Clausius's uncompensated heat with affinity. Because of its generality and precision, this definition easily permitted us to extend to irreversible transformations of real systems the classical formulas of perfect gases." (De Donder 1925b, p. 1). He thus reviewed his work by presenting the concept of specific affinity and some new developments. The communication is written with the same pedagogical style that characterized *Lectures I* and *Lectures II*. Hence, he conceived his communication as a presentation for a new audience. In a footnote (Fig. 9), De Donder referred to the discussion of André Job's report published in the proceedings of the first Solvay Chemistry Council (De Donder 1925b, p. 10)[20], which took place in Brussels in 1922. More precisely, De Donder referred to two different objections that occurred during the discussion, one raised by the Swiss chemist Alfred Berthoud and the other by the Belgian colleague Frédéric Swarts. De Donder did not participate in the conference of 1922, hence, he should have completed his communication after the publication of the volume, i.e. the end of March 1925. This proves that De Donder conceived it to emphasize the importance of his approach to solve the questions addressed during the first Chemistry Council in 1922.

---

[17] English translation: "To the Members of the Solvay International Chemistry Congress, April 1925. see page 21".

[18] English translation: "To Mr Charles Lefébure Acknowledgments and tributes".

[19] We are aware at least of two of them. One is conserved at the Solvay Archives and the other at the *Réserve Précieuse* archive of the ULB.

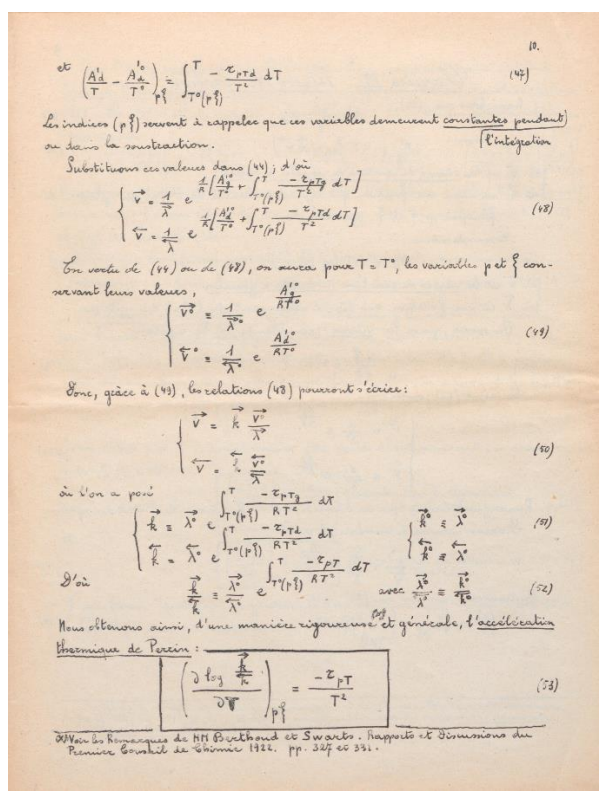

**Fig. 9** The footnote of the Solvay communication referring to the proceedings of the first Solvay Chemistry Council

De Donder's communication for the second Chemistry Council is not dated, but it can be argued that it was conceived between April 4 and the beginning of the Council for the following reasons. First, as already said, on April

---

[20] In the digitalized version of this communication on the website of the Solvay Science Project, the footnote is not visible.

4 De Donder was at the presentation of the proceedings of the first Chemistry Council and he asked Dony-Hénault to present a communication. Second, the second Chemistry Council took place in Brussels on April 16–23, while (De Donder 1925a) containing some of the material discussed in the communication to the Council, was published on April 27. Hence, by preparing his paper, De Donder decided to discuss some open questions of the first Council. We notice that in (De Donder 1925a) he did not make any reference to the Chemistry Council, because, as we shall see in section 5, he supposed that his communication could be inserted into the proceedings of the second Council.

Let us now consider the connection suggested by De Donder between his approach and the discussions of the first Solvay Chemistry Council. The conference, held in Brussels in 1922, was devoted to the discussion of five topical issues. One of the themes was encoded in the title of the report prepared by Job: *Chemical Mobility*. The report was read by Jean Perrin because, due to a severe illness, Job was absent. The explicit aim of Job's report was to place under the scrutiny of chemists the possibility of advancements in Chemistry through the cross-fertilization with Physics. He started by noting that in the last 30 years, there had been a huge increase in the number of papers investigating the topic of chemical kinetics. In this context, he emphasized: "The moment is quite well-chosen to reflect and to deliberate, because new suggestions come to us, inspired by the physicists. More precisely, we will focus on the *radiation hypothesis* [hypothèse radiochimique]. It is now the time to expose it to the chemists and let them discuss it." (Proceedings 1925, p. 284). A brief explanation of this hypothesis is needed to understand De Donder's contribution.

The chemists had turned their attention from the end of the 19$^{th}$ century to the study of the velocity of reactions and its connection with the probability of reaction. Special attention was dedicated to unimolecular reactions, i.e. those reactions where the transformation of a reactant molecule $A$ into a product molecule $A'$ occurs in a single chemical step. In this context, in 1919 Jean Perrin had proposed that unimolecular chemical reactions could be explained in terms of blackbody radiation from the reaction vessel. This was the so-called radiation hypothesis and its impact on the history of chemical kinetics is well described in King and Laidler (1984). Perrin's radiation hypothesis was based on two conjectures: first, the reacting substance has to absorb radiation at the frequency required for activation and, second, the radiation density has to be high enough to supply the energy for activation. For unimolecular reactions, Job considered the existence of an intermediate state $I$ and defined the rate of reaction $k$ as the product of two different quantities. The first, which would depend on the temperature $T$, was the ratio between the concentration of the molecule $A$ and the intermediate state $I$. The second, which would not depend on the temperature, is the number of molecules per unit of time, relative to the concentration of the intermediate state $I$, which transforms into the final molecule $A'$. In this context, Job discussed the thermal acceleration introduced by Perrin, namely $\frac{d \ln(k)}{dT}$, and noticed how Svante Arrhenius obtained that $k$ is proportional to an exponential factor, also known as the Arrhenius activity formula (Nye 1993, p. 123).

During the discussions at the first Solvay Chemistry Council, Berthoud noticed that, given a reversible unimolecular reaction, the radiation hypothesis would imply that the thermal acceleration of the ratio of the two equilibrium constants, i.e. forward over reverse reaction constant $\frac{\vec{k}}{\overleftarrow{k}}$, should be equal to $\frac{q}{RT^2}$, where $R$ is the universal constant of gases, $T$ is the absolute temperature, and $q$ should be the heat reaction. Berthoud criticized the radiation hypothesis because it cannot explain the fact that $q$ could vary with the temperature $T$. In the same discussion, Swarts insisted on the fact that both the forward and the reverse reaction constants should depend on temperature, especially for reversible processes, and emphasized that it was problematic to reconcile the radiation hypothesis with Gibbs-van 't Hoff's equation, in the presence of the intermediate state considered by Job. By referring to these critics in his communication written for the second Solvay Council, De Donder derived the formula for thermal acceleration involving the reaction constants $k$ by using thermodynamical arguments and an Arrhenius-like ansatz for the velocity of the reaction. But his formula connected the velocity of reaction, i.e. the ratio of the reaction constants $\frac{\vec{k}}{\overleftarrow{k}}$, with the concept of chemical affinity.

The *Solvay communication* contains also other improvements of the concept of chemical affinity. In the communication, De Donder gave proof that the specific chemical affinity is a function of state and he explained in detail the relationship between the specific affinity and all the thermodynamic potentials, including Clausius internal energy, Kamerlingh's Onnes enthalpy, Helmholtz's free energy and Duhem's thermodynamic potential (De Donder 1925b, p. 4). This last potential is Gibbs free energy: here De Donder wrote for the first time that $\mathcal{A} = -\left(\frac{\partial G}{\partial \xi}\right)_{pT}$. In the *Solvay communication*, De Donder changed the symbol for the extent of reaction from µ to ξ. The motivation will be explained at the end of this section. De Donder distinguished between the velocity of a direct and an inverse chemical reaction, where now the former is the one running from left to right, and he presented a result that he claimed to be connected with the discussions of the first Solvay Chemistry Council. Regarding the direction of the chemical reaction, De Donder described for the first time the meaning of the variable that represents the degree of advancement by drawing it as an axis going from the left to the right. From this perspective, the auxiliary variable increases when the reaction goes in the same direction and, as we have anticipated in the preceding section, the Lavoisier principle was consequently represented as $0 = \sum_{\gamma,a} v_\gamma^a M_\gamma^a \frac{\partial G}{\partial m_\gamma^a}$ to emphasize that the stoichiometric numbers are positive for the constituents on the right-hand side of the reaction.

In the proceedings of the first and the second Solvay Conferences, the term affinity refers always to electronic affinity, which is different from Do Donder's concept. The importance of De Donder's result was not recognized at the time: in his approach, the reaction heat is not necessarily constant with temperature and his formula permitted him the introduction of the concept of irreversibility into chemical kinetics. There is no trace of De Donder's influence in the proceedings of the second Chemistry Council, but, as already said, even if he was not allowed to present his work, he should have been invited to the banquet where he should have had the possibility of presenting it. Hence, we investigated whether he discussed his approach with the people that attended the conference. We don't know if the copies prepared by De Donder were distributed to the participants, but we know that he was officially invited to the banquet offered by Solvay. Indeed, his name is on the list of participants (Fig. 10). The banquet took place at the *Taverne Royale* in Galerie du Roy (Passage St. Hubert) on April 22. The definitive menu (Menu 1925) is reported in Fig. 10, but De Donder did not participate. As we argued, he had put some effort to present his point of view, hence, we tried to understand what happened.

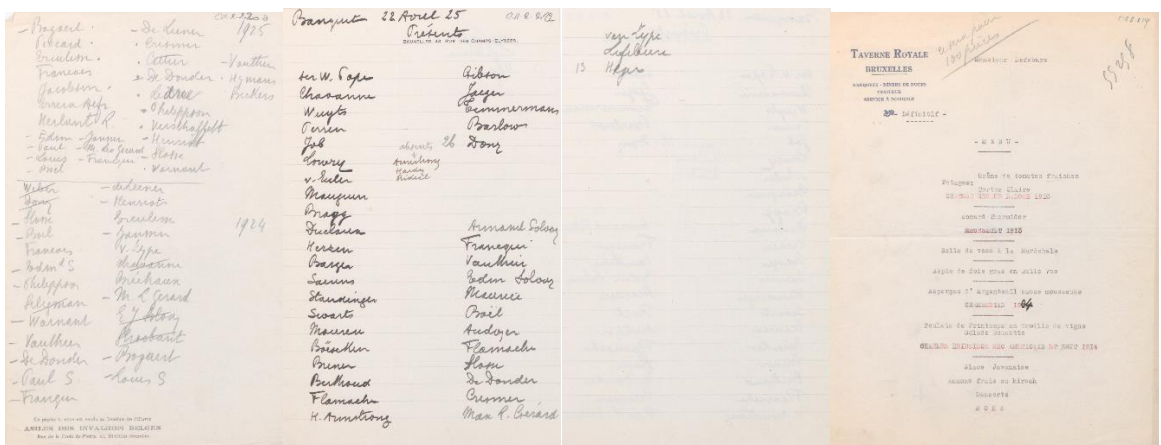

**Fig. 10** List of participants and *Taverne Royale*'s menu. The list on the left compares the participants to the banquet of the first Chemistry Council (1922) with the participants of the second conference (1925) De Donder's name is present. The two central images contain the *supposed* full list again with De Donder's name. On the right the dinner menu

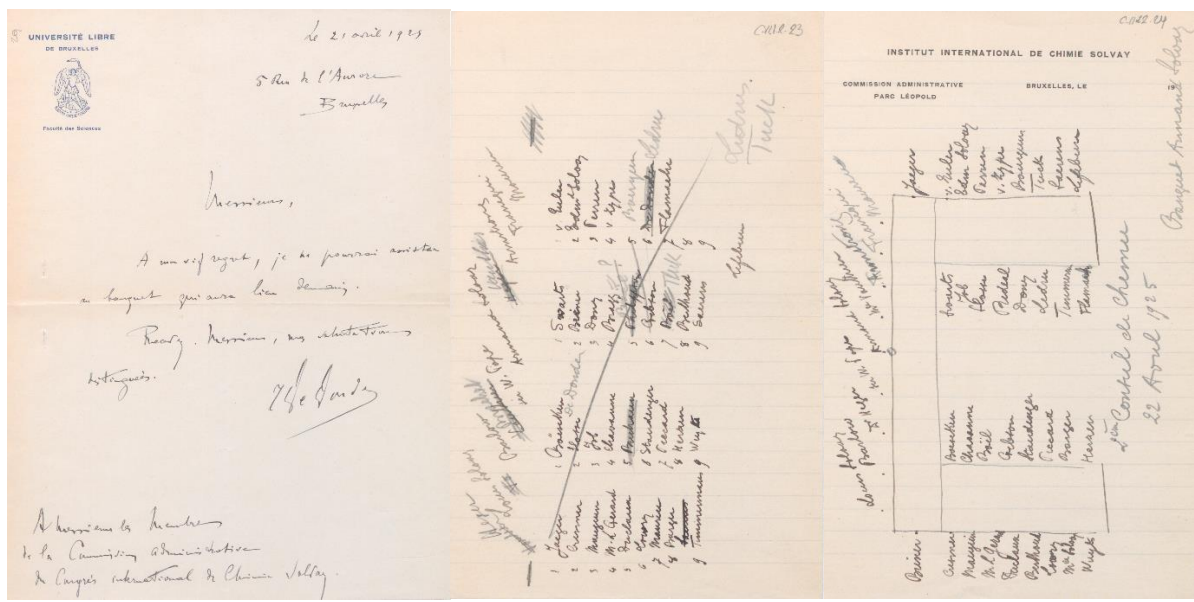

**Fig. 11** De Donder's letter and notes of the banquet. On the left, De Donder's letter announcing that he will not be present. In the center, De Donder's name is cancelled with a pencil and the name *Ledru* is written on its right. On the right, De Donder's name is not present. There is Ledru's name instead.

Among the documents conserved at the Solvay Archive, there are three notes (Notes 1925) for the seating arrangements at the dinner table (Fig. 11): in the first one, De Donder's name is present, but it is cancelled with a pencil and the name *Ledru* is written on the right. In the final annotation with the seating arrangements, the name of De Donder has been replaced by Ledru. In a letter dated April 21, i.e. the day before the banquet, De Donder communicated his inability to participate (De Donder 1925c), without any explanation (Fig. 11). Even if we do not have direct evidence of what prevented De Donder from participating, we can argue that his decision was connected with the health of his wife. As we already said, on March 16, 1925, De Donder received a letter from his friend and colleague Heymans. From this letter, we are informed that when De Donder wrote to Heymans, i.e. on February, 17, his wife had some health problems and that her health was not improving as rapidly as De Donder hoped (Fig. 6). The dinner was scheduled only two months after De Donder's letter to Heymans, hence it is plausible that De Donder hoped to be present at the *Taverne Royale*, maybe with her[21], but because of her health problems he waited until the day before the dinner to make his final decision.. Therefore, even if his communication has been distributed to the participants, the content was not analyzed either during the conference, because it was not inserted in the published proceedings, or during the dinner, because of his absence. His mathematical approach was distant from the dominant attitude in Chemistry at the time, which was primarily based on the importance of experimental evidence rather than theoretical activity. This fact and the evidence of his absence to the dinner clarify why there is no trace of his work in the proceedings.

Even if the *Solvay Communication* was not inserted in the proceeding at the conference or discussed with the Chemistry Council, it represented a key moment in the evolution of the concept of affinity in De Donder's work. As already said, in this unpublished communication De Donder elaborated some improvements that will be published in (De Donder 1925a). One of these improvements can be understand by clarifying why De Donder changed the symbol representing the extent of reaction from $\mu$ to $\xi$. He did not make a mere graphical change: the two variables are related by a minus sign. The relation $\xi = -\mu$ corresponded to a change in direction of the axis representing the system evolving in time. Hence, De Donder defined a new extent of reaction and for a new definition a new symbol was needed. The introduction of the new symbol was equivalent to the introduction of a new perspective. Indeed, also the formal representation of the chemical reactions changed. In 1920, De Donder represented the law of conservation of mass as follows, i.e. $\sum_\gamma \nu_\gamma M_\gamma = 0$, where the index $\gamma$ labelled the constituents of the reaction, $\nu_\gamma$ are the stoichiometric coefficients and $M_\gamma$ are the molar masses, because he adapted to the convention used by the chemists at the time (De Donder 1920c, p. 104) who assigned positive stoichiometric numbers to the products of the reaction, i.e. the species appearing on the right-hand side of a chemical reaction[22]. From 1925, he would instead write the law of conservation of mass as $0 = \sum_\gamma \nu_\gamma M_\gamma$. The new writing underlined that he started to adopt an opposite convention: the positive numbers were now assigned to the reactants and hence a positive increment of the extent of reaction corresponded to the advancement from the left to the right, like for a usual Cartesian horizontal axis.

---

[21] In 1927, at the dinner organized for the fifth Solvay Physics Council, De Donder was invited with his wife.

[22] For example, with this old convention the stoichiometric coefficient of the reaction $N_2 + 3H_2 \rightarrow 2NH_3$ are $\nu_1 = 1$, $\nu_2 = 3$ and $\nu_3 = -2$.

## 5 De Donder's *Ave Maria* (1932)

De Donder's unpublished communication for the Solvay conference is important also for another reason. In another footnote, he wanted to inform the participants of the conference that van Lerberghe was also working on the same topic and that he had obtained some results concerning the relationship between the velocity of a chemical reaction and the specific affinity. The first work published by van Lerberghe on De Donder's affinity appeared only in June (Van Lerberghe 1925), i.e. after the second Solvay Chemistry Council. The results quoted by De Donder would be published by van Lerberghe the next year (Van Lerberghe 1926b, p. 526). As recollected by De Donder, remembering his disciple in 1940, after van Lerberghe's death that occurred at the front during the Second World War, "he embraced the new point of view of the affinity and the velocity of reaction, which opened the gate to some discoveries where he as author affirmed his vigorous personality and an incontestable mastery" (De Donder 1940). The correspondence between van Lerberghe and De Donder confirms that the two Belgian scientists constantly discussed the development of chemical affinity (Fig. 12).

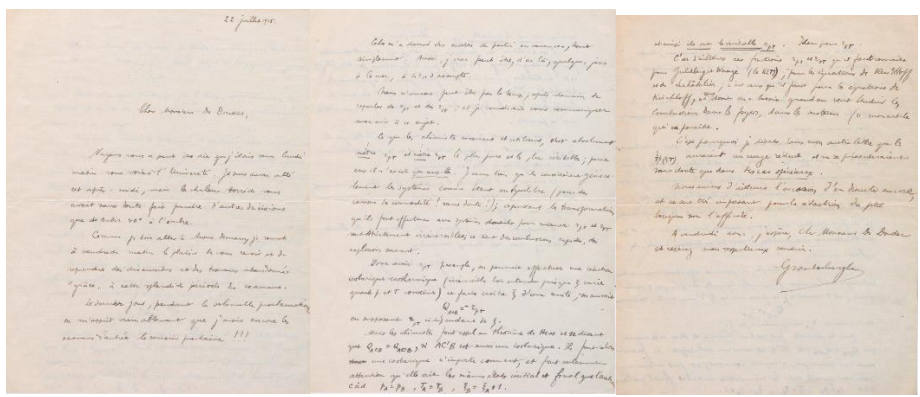

**Fig. 12** Van Lerberghe's letter discussing the heat reaction (Van Lerberghe 1925)

From the same correspondence, it emerges that De Donder was disappointed with the Chemistry Council. The year after the conference, in 1926, De Donder and his wife finally travelled to Boston, in the United States, to visit Heymans at MIT, where De Donder had been invited to give a series of lectures on general relativity. This visit was important to him because it stimulated his work aiming to merge wave mechanics with general relativity (Peruzzi and Rocci 2022). He did not lecture on chemical affinity, but we have two pieces of evidence pointing to the fact that he also discussed this topic. As we have already said, Heymans's letter suggests that Heymans and Debye were anxious to know De Donder's point of view. Furthermore, on the back-side of the postcard advertising De Donder's lectures at MIT, there are some confused annotations written in pencil by De Donder that suggest a rough explanation of the subject (Fig. 13). The correspondence between De Donder's handwriting and the annotations of Fig. 13 can be confirmed by comparing the squared formula, which reads $dQ = C_{p\xi} + h_{T\xi}dp$. The variable $p$, which indicates the pressure, can be compared with the letter[23] *p* on the world *page* of Fig. 8: the two letters seem to match. Hence, we can suppose that the handwriting of Fig. 13 is that of De Donder. We notice that the squared formula indicates the differential of the usual function $Q$, which denotes the heat, and that De Donder added out of the square the differential coming from the generalization of the function $Q$, obtained by adding the extent of reaction as a new variable. The capitalized letters $\mathcal{A}$ indicate the specific affinity and it is the same character appearing in the *Solvay Communication* (De Donder 1925a). In Fig. 13 both the equilibrium and the out-of-equilibrium case are considered, i.e. $\mathcal{A} = 0$ and $\mathcal{A} \neq 0$ respectively. As a confirmation of the fact that De Donder was invited to MIT to discuss both general relativity and his work on affinity, in the archive of the ULB, there is an annotation in French written by Jean Paul Emile Bosquet, another disciple of De Donder (Fig. 13). A translation of Bosquet's annotation reads: "De Donder was on his way to MIT […] he was invited to lecture on the theory of relativity as well on electricity, affinity…".

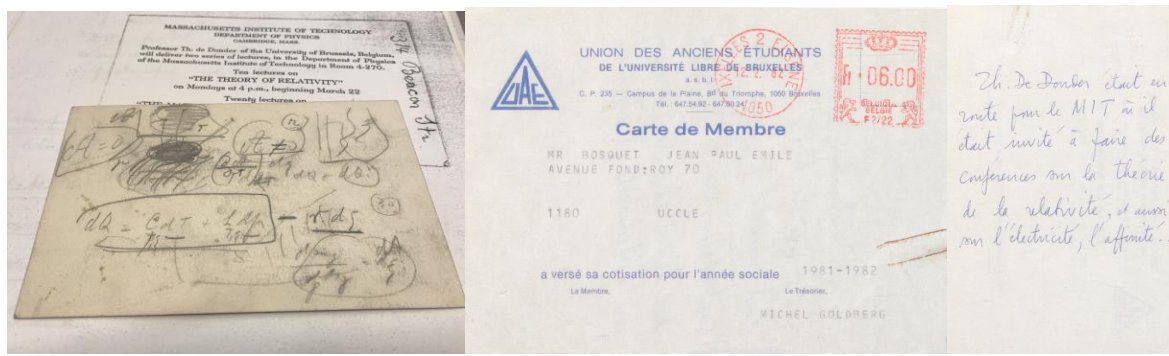

**Fig. 13** On the left to the right. De Donder'a annotations, Bosquet's postcard and its back with Bosquet's annotation

During his visit to MIT, De Donder continued to discuss chemical affinity and kinetics with van Lerberghe (Van Lerberghe 1926a), who was going to publish another paper on these subjects (Van Lerberghe 1926b),. In a letter dated May 1, 1926 (Fig. 14), van Lerberghe informed De Donder that he had in his hands, even if for a very short period of time, the proceedings of the second Chemistry Council and that he had searched for De Donder's note. From this, we can infer that De Donder hoped that his note could be published, because he wanted to share his work with the international community of chemists. Indeed, van Lerberghe said that he was surprised by this omission and he tried to ask Dony-Henault for some explanation. He finally added ironically that Dony-Henault was the least informed person to ask. After this episode, De Donder would be more involved in the Solvay Councils, but in the context of Physics conferences. Indeed, he would be invited to all of them until 1948 and he would be a member of the scientific committee from 1929 (Mawhin 2012, p. 65). As far as we are aware, he never tried to repeat the experience with chemists. The concept of affinity appeared in the context of the Physics conferences. At the end of the proceedings of the seventh Solvay Physics Council, there is a note without discussion communicated by De Donder[24] (Solvay Proceedings 1933, p. 340). Chemical affinity reappeared after his death in 1980, at the 17th Solvay Chemistry Council entitled *Advances in chemical physics*, because his disciple Prigogine celebrated De Donder's work and its developments when he introduced the rate of entropy production for irreversible processes[25], a modern point of view on Clausius's uncompensated heat.

---

[23] The letter *e* cannot be a good term of comparison, because it is different in the two handwritings of Fig. 8, maybe because one of the two was written in a fast way.

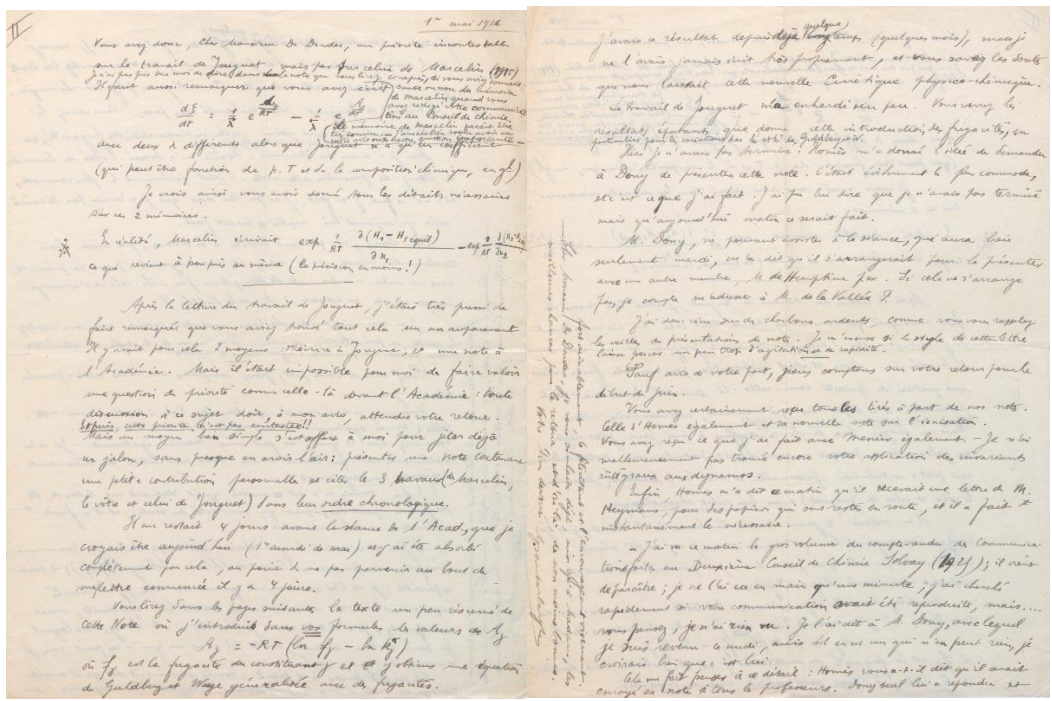

**Fig. 14** Letter of van Lerberghe to De Donder (1926). On the right, above we recognize the symbol for the specific chemical affinity introduced by De Donder. On the left, at the bottom, "Conceil de Chimie Solvay (1925)" can be recognized

In his homage to De Donder, van den Dungen underlined that the Belgian mathematician published periodically some developments on the concept of affinity. As already said, the first papers are dated 1922 and 1925. He added some new results in 1927, 1931 and 1933. De Donder published the content of the *Solvay communication* in his future papers. For example, the generalization of Perrin's thermal acceleration (van 't Hoff equation) is presented in 1927 in his work entitled *L'Affinitè* (De Donder 1927). We recall that at the end of section 2 we have underlined that at the time the chemist community was unprepared to appreciate De Donder's formal approach. How was his work on affinity received? A preliminary analysis shows that De Donder's approach did not receive much attention from the community of chemists until the end of the Second World War. Let us briefly consider what happened in Belgium and the context of the English-speaking community[26]. In Belgium, his disciples started to develop De Donder's ideas after 1925. At the beginning of this section, we have briefly explained the role played by van Lerberghe. Another person that published some results on this topic is Georges Homès, a future professor at the ULB and the University of Mons. He started his career by publishing a paper at the end of 1925 on the equilibrium of physicochemical systems, where he developed De Donder's affinity (Homès 1925). The international community, especially the English-speaking part, seemed not to consider De Donder's work. As far as we are aware, the first quotation of De Donder's work on affinity appeared in 1930 in the journal *Physical Review* (Beattie 1930). This lack of interest from the English-speaking community would also be noticed by one of De Donder's future collaborators, Peter van Rysselberghe in 1935, when he embarked on the translation into English of the work of De Donder and his collaborators. Except for De Donder's disciples, we found no papers until the mid 1930s investigating the mathematical concept introduced by De Donder either in the *Bulletin* of the Belgian Royal Academy or in the *Comptes Rendus* of the French Academy of Sciences, the two journals where De Donder published his work. With the end of the 1930s, the advent of the Second World War considerably slowed the research, even if De Donder never stopped either working or teaching, like in the preceding global conflict. As anticipated, De Donder's and his disciples' work was generally ignored until the end of the war. A possible explanation is as follows. According to the American historian Henry Leicester, in much of the English-speaking world, affinity was not perceived as a fundamental concept. Instead, the scientists based the study of chemical thermodynamics on Gibbs's free energy. The reason for this fact can be ascribed to the influence of Gilbert N. Lewis and Merle Randall's textbook[27] (Jensen 2015, p. 136; Kragh and Weininger 1996; Leicester 1951), which "became the bible in the field for chemistry students" (Devine 1983, p. 333), and insisted explicitly on the necessity of the transition from the idea of affinity to the concept of free energy. Indeed, after having reviewed the "modern stage of thermodynamics" (Lewis and Randall, p. 5) at that time, Lewis and Randall noticed that the research area was in its third stage, "characterized by the design of more specific thermodynamic methods and their application to particular chemical processes, together with a systematic accumulation and utilization of the data of chemical thermodynamics." (Lewis and Randall, p. 5). The authors underlined: "In the United States there has been a widespread interest in the chemical applications of thermodynamics" (Lewis and Randall, p. 6). In this context, i.e. by devoting themselves to questions of applied thermodynamics, the two authors pointed out that "the work of Thomsen, Berthelot and others has given us a great mass of thermochemical data of all grades of accuracy" (Lewis and Randall, p. 97), but they pointed out that "the hope of these investigators that the results of their labors would give a direct measure of chemical affinity has proved to be a vain one" (Lewis and Randall, p. 97). Finally, the two authors emphasized that "the study of free energy affords the only true measure of chemical affinity." (Lewis and Randall, p. 584). Lewis and Randall's philosophy influenced the reception of De Donder's work based on the concept of affinity.

---

[24] De Donder tried to generalize the concept of affinity in the context of what he called *relativistic thermodynamics*. We postpone the discussion of this topic to a future work.

[25] For an explanation of this concept see (Kondepudi and Prigogine 2015, p. 131).

As we have underlined, the two innovative concepts introduced by De Donder in physical chemistry are the extent of reaction and chemical affinity. Jensen emphasized how the IUPAC accorded official recognition to De Donder's extent of reaction in the 1980s (Jensen 2015, p. 171), but we found that both extent of reaction and chemical affinity already appeared in the 1973 Edition of the *Manual of symbols and terminology for physicochemical quantities and units* (adopted by the IUPAC Council at Cortina d'Ampezzo in 1969), which would be called the *Green Book*, (Paul and Whiffen 1973, p. 11 and p. 32). The two concepts are related through the so-called De Donder's inequality. For a closed system undergoing a single chemical reaction, the inequality describes the time development of the chemical process as follows, namely $\mathcal{A} \cdot v \geq 0$, where the overall rate $v$ is the algebraic sum of the forward and the reverse reaction velocities. For $m$ stoichiometrically independent reactions, De Donder's inequality is formulated as follows: $\sum_{i=1}^{m} \mathcal{A}_i \cdot v_i \geq 0$. De Donder was especially attached to this formula. The importance of De Donder's achievement has been emphasized by his disciple Géhéniau, who underlined that De Donder's inequality "is the translation in the mathematical language of Le Chatelier's principle" (Géhéniau 1968, p. 181). To understand Géhéniau's remark, let us recall the statement known as the principle of Le Chatelier as it is presented in Prigogine's textbook:

> "When a system is perturbed from its state of equilibrium, it will relax to a new state of equilibrium. Le Chatelier and Braun noted in 1888 that a simple principle may be used to predict the direction of the response to a perturbation from equilibrium. Le Chatelier stated this principle thus: Any system in chemical equilibrium undergoes, as a result of a variation in one of the factors governing the equilibrium, a compensating change in a direction such that, had this change occurred alone it would have produced a variation of the factors considered in the opposite direction." (Kondepudi and Prigogine 2015, p. 242)

---

[26] We leave aside the study of the impact in Europe.

[27] For a review of the textbook and a tribute to its authors see (Jensen 2015, p. 129).

De Donder inequality permitted to identify the change in direction with the sign of affinity, which had finally a precise mathematical form. De Donder expressed more firmly the same conviction by stating that his inequality gives "the general and *correct* form" [emphasis added] (Géhéniau 1968, p. 181; De Donder 1936, p. 97) of Le Chatelier's principle. Quoting De Donder's words, Géhéniau continued as follows: "The inequality $\mathcal{A} \cdot v \geq 0$ gives in all cases the exact sign [sense] of the reaction velocity resulting from the perturbation. It also shows, without ambiguity, the conditions under which the reaction may bring about a moderation of this perturbation."[28] (De Donder 1936, p. 104; Géhéniau 1968, p. 181). This inequality was a sort of *closing the loop* for De Donder, because it showed once again how the state of equilibrium is represented by a minimum of a function, in this case, the affinity, the same idea that he had started to investigate with the creation of his *Lecture I* around the end of 1910s.

Even if De Donder's inequality easily follows from the definition of specific affinity and the second law of thermodynamics, he wrote it explicitly for the first time in 1931, in its second general form, and he discussed it more extensively in 1932. At the time of the communication for the Chemistry Council, in 1925, De Donder had all the elements to write it. Indeed, at the beginning of the *Solvay communication*, he noticed that the uncompensated heat is a non-decreasing function of time, namely $\frac{dQ^n}{dt} \geq 0$. This was the first time that De Donder explicitly introduced the time variable, through the use of the extent of reaction, for describing the dynamics of the systems. In the communication, De Donder was interested in the graphical meaning of this inequality for a system undergoing a cyclic transformation and, on p. 2 of the *Solvay Communication*, he drew three diagrams showing how the area enclosed by the curve representing the transformation in the Clapeyron plane should be smaller than the area representing the same process of the entropy/temperature diagram. In the succeeding section, by discussing mono-phasic chemical reactions, De Donder considered the difference between the time derivative of the uncompensated heat and the derivative with respect to the degree of advancement of the reaction. At this point, he briefly mentioned that the time derivative can be rewritten as follows applying the theorem on functions of functions, namely $\frac{dQ^n}{dt} = \frac{dQ^n}{d\xi}\frac{d\xi}{dt} = \mathcal{A} \cdot v$ (De Donder 1925b, p. 4). By merging these two statements, he should have been able to discover his inequality, but it seems that he had not yet recognized the connection with Le Chatelier's principle. A decisive step in this direction took place in 1932 when he explicitly discussed the "power of the system" (De Donder 1932, p. 881) undergoing chemical reactions, namely $P = \frac{dQ^n}{dt}$. In this paper, De Donder gave the net outflow of uncompensated heat a specific name and defined his inequality as "an extremely remarkable result for chemically active physical-chemical systems" (De Donder 1932, p. 882). He emphasized the general validity of the result by underlining that "it is independent of the conditions to which this system is subjected during the reaction from $t$ to $t + dt$" (De Donder 1932, p. 883). For De Donder, his inequality was one of the main outcomes of his approach because it connects the concept of affinity with the direction of the reaction. This is testified especially in his books on chemical affinity, both the original French and the English version, where he referred to it as "the *fundamental* inequality"[emphasis added] (De Donder 1936, p. IX). He expressed similar appreciation also in conferences that were open to the public, as when he was asked to celebrate the annual public meeting of the Science Class of the Royal Belgian Academy (Fig. 15), at the end of 1937, as director of the Class.

---

[28] The statement quoted is the (official) English translation provided by van Rysselberghe for De Donder's book and it is the same statement quoted by Géhéniau (in French).

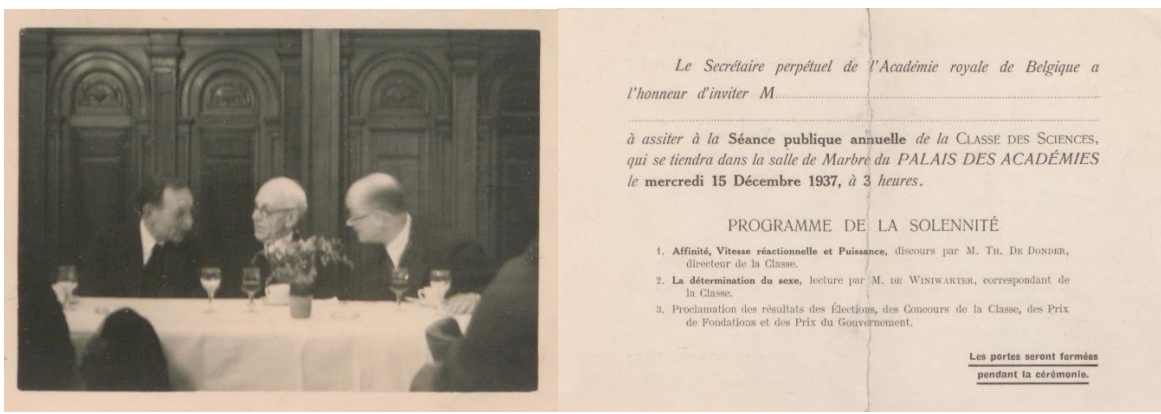

**Fig. 15** On the left, De Donder is sitting at the center during the annual public meeting of the Science Class of the Royal Belgian Academy (1937). On the right, the program of the celebration

In his speech entitled *Affinity, the velocity of reaction and power*, De Donder emphasized the importance of his inequality because of its general validity and its simplicity, and once again he pointed out the role of thermodynamics: "This link, so simple, between the speed with which a reaction run and his ultimate cause, essentially chemical, the *Affinity*, has been sensed since time immemorial. Thermodynamics alone could have shed light and provided precision." (De Donder 1937, p. 979). His approach based on the methods of mathematical physics, which contributed to isolating him from the community of chemists, was to him the key to understanding the nature of irreversible processes: "Once again, mathematical chemistry proves indispensable to understand quantitatively and from a general point of view the links between affinity and reaction energies." (De Donder 1937, p. 980). In the same period, the future electrochemist Marcel Pourbaix, who graduated from the University of Brussels in 1927 and was a PhD candidate during the years before WWII, extended De Donder's inequality to an electrochemical reaction following a conversation with him. We know from Pourbaix's later recollections that De Donder was so attached to his inequality that he called it his "Ave Maria" (Pourbaix 1984, p. 89). Once again, we are informed about how much De Donder was impressed by the importance and the fruitfulness of this achievement: "In 1936, I carried out several laboratory experiments which showed that this [electrochemical extension of "Ave Maria"] was correct and useful. When De Donder heard about this, he jumped for joy (he was still like a child, despite his 64 years of age) saying, *The good Lord is the most simple being in the Universe*." (Pourbaix 1984, p. 89).

**6 Summary and conclusions**

With an unusual approach from the point of view of chemists of the time, De Donder opened the door to a revolution in thermodynamics and chemical kinetics. The methods of mathematical physics or, as he liked to call it, *mathematical chemistry* permitted De Donder to give the concept of affinity a precise formulation. He began to rethink the approach to thermodynamics at the beginning of his teaching career by lecturing on mathematical physics at the University of Brussels. He felt comfortable with mathematical methods introduced by Gibbs and Duhem, but he was disturbed by the asymmetry between reversible and irreversible processes. By focusing on the extension to real processes of the thermodynamic potentials, he introduced the extent of reaction and focused on the displacement from the equilibrium. After having discovered the role of Clausius's uncompensated heat and after having formalized it, he proposed to identify it with the concept of chemical affinity in 1922. He finally identified the connection between chemical affinity and Gibbs free energy, formalizing the old idea connecting affinity with the concept of attractive force in 1925. In the same year, the extent of reaction started to point to the right. He changed the formal representation of chemical reactions and he gave proof of the fact that chemical affinity is a new function of state. His efforts in calling the attention of the international community of chemists to his approach produced the mentioned development of the early concepts. He tried to interact with the participants of the second Solvay Chemistry Council, but his work did not receive attention, maybe also because of his inability to participate to the banquet celebrating the end of the conference. Despite this fact, his pedagogical attitude and his efforts in teaching his developments formed many disciples and they became his close collaborators. The list mentioned in this paper is not exhaustive. Using a chronological order, we briefly pointed out how van Lerberghe, Homés and van Rysselberghe contributed to the development of all the concepts related to chemical affinity. These collaborators, with Géhéniau, Bosquet, van den Dungen, Glansdorff, Pourbaix and Prigogine who contributed with their remembrance to reconstruct De Donder's figure, are part of the *Brussels school of thermodynamics* (Glansdorff et al. 1987, p. 7; Kondepudi and Prigogine 2015, p. xxi), a research group that formed during the period considered in this paper. The school developed De Donder's concepts and opened the road to modern thermodynamics, where, as De Donder wanted, both equilibrium and out-of-equilibrium processes are treated on the same footing. We have therefore traced the path for reconstructing the history of this research group that expressed a Nobel prize in physics.

Both De Donder's last representation of the degree of advancement and chemical affinity concepts are now accepted in physical chemistry. Among all his mathematical outcomes in this framework, De Donder obtained a simple and, in his own words, *fundamental* inequality involving both the affinity and the velocity of the reaction. It encoded the meaning of affinity, by describing how it rules in which direction the chemical reaction must run. The inequality connected the thermodynamics with the realm of chemical kinetics by allowing the concept of irreversibility to connect the two research areas and, as De Donder said, it contained in its simplicity the beauty of the Universe. The concept of uncompensated heat is no longer part of the vocabulary of the chemists because De Donder's disciples, starting from his achievements, had exchanged it for the concept of entropy production. But the extent of the reaction and his chemical affinity survived. The work and the impact of the Belgian mathematician are well described and summarized by the words of Marcel Pourbaix: "De Donder was a kind, gentle pioneer and revolutionist during an era when a majority of eminent chemists wanted chemistry to remain essentially experimental and reproached the theorists who favoured the invasion of Mathematics in a science which could, if necessary, not involve any mathematics." (Pourbaix 1984, p. 93). Despite its strong empirical character, chemistry benefited De Donder's formal approach. Using again his words: "Mathematical chemistry proved indispensable to understanding quantitatively and from a general point of view the links between affinity and reaction energies" (De Donder 1937, p. 978).

**Acknowledgements**

The author benefited the help of Marina Solvay, and his colleagues Yoanna Alexiou and Nicolas Brunmayr in understanding some handwritten French words and from the support of the archivists of the Université Libre de Bruxelles (ULB) for the digitalization of the documents. The author is also grateful to Franklin Lambert for the interesting discussions, Brigitte van Tiggelen, Kenneth Bertram and Jan Danckaert for their contribution to the Solvay Science Project and his colleague Catherine Judson for revising the English language and for her fruitful suggestions.

*Funding*: This work was supported by the Fund for natural sciences in society at the Vrije Universiteit Brussel (VUB)